\documentclass[aps,prb,twocolumn,twoside,superscriptaddress]{revtex4-2}
\usepackage[version=3]{mhchem} 
\usepackage{amsmath}
\usepackage{amssymb}
\usepackage{amsthm}
\usepackage{amsfonts}
\usepackage{listings}
\usepackage{longtable}
\usepackage{enumerate}
\usepackage{latexsym}
\usepackage{color}
\usepackage{setspace} 
\usepackage{blindtext}
\usepackage{dsfont}
\usepackage{mathrsfs}
\usepackage{array,etoolbox}
\usepackage{bm}
\usepackage{hyperref}
\usepackage{graphicx}
\usepackage{subfigure}
\usepackage{psfrag}
\usepackage{changes}
\usepackage{ulem}

\begin{document}

\newcommand{\TODO}[1]{\textcolor{red}{#1}}
\newcommand{\super}[1]{\ensuremath{^{\mathrm{#1}}}}

\title{Quantum simulation of carbon capture in periodic metal-organic frameworks} 
\setcounter{page}{1}

\author{Dario Rocca}
\affiliation{QC Ware Corporation,
Palo Alto, California 94301, USA}

\author{J\'er\^ome F. Gonthier}
\affiliation{QC Ware Corporation,
Palo Alto, California 94301, USA}

\author{Joshua Levin}
\email{joshua.levin@qcware.com}
\affiliation{QC Ware Corporation,
Palo Alto, California 94301, USA}

\author{Tobias Sch\"afer}
\affiliation{Institute for Theoretical Physics, TU Wien, Wiedner Hauptstra\ss e 8–10/136, 1040 Vienna, Austria}

\author{Andreas Gr\"uneis}
\affiliation{Institute for Theoretical Physics, TU Wien, Wiedner Hauptstra\ss e 8–10/136, 1040 Vienna, Austria}

\author{Byeol Kang}
\email{bkang@posco-inc.com}
\affiliation{POSCO Holdings,
Pohang, Republic of Korea}

\author{Hong Woo Lee}
\affiliation{POSCO Holdings,
Pohang, Republic of Korea}

\begin{abstract}
Carbon capture is vital for decarbonizing heavy industries such as steel and chemicals. Metal-organic frameworks (MOFs), with their high surface area and structural tunability, are promising materials for CO$_2$ capture. This study focuses on Fe-MOF-74, a magnetic Mott insulator with exposed metal sites that enhance CO$_2$ adsorption. Its strongly correlated electronic structure challenges standard DFT methods, which often yield inconsistent predictions. We initially benchmark adsorption energies using various DFT functionals, revealing substantial variability and underscoring the need for more accurate approaches, such as those provided by quantum computing. However, practical quantum algorithms are far less
established for simulations of periodic materials, particularly when the plane-wave basis set—often comprising tens of thousands of basis vectors—is used. 
To address this, we employ an active space reduction strategy
based on Wannier functions and natural orbital selection. Localized orbitals around the adsorption site are identified, and MP2 natural orbitals are used to improve convergence of correlation energies. Adsorption energies are then computed using a quantum number-preserving ansatz within the variational quantum eigensolver framework. In addition to classical simulations, we conduct quantum experiments using the sample-based quantum diagonalization method. Although current hardware limits the size of feasible simulations, our approach offers a more efficient and scalable path forward. These results advance the applicability of quantum algorithms to realistic models of carbon capture and periodic materials more broadly.
\end{abstract}

\maketitle

\section{Introduction}

Carbon capture technologies are crucial for the decarbonization of energy-intensive industries such as steel and chemical manufacturing. Among emerging materials, metal-organic frameworks (MOFs) stand out due to their exceptional structural tunability and high surface areas, making them prime candidates for efficient CO$_2$ capture \cite{sumida2012carbon,ding2019carbon}. However, accurately modeling CO$_2$ adsorption in periodic solids remains a major challenge. Traditional density functional theory (DFT) methods often struggle to provide a reliable description of these complex systems where van der Waals forces and strong electronic correlation play an important role \cite{klimevs2012perspective,anisimov1997first}. Given their ability to naturally incorporate many-body effects and handle large computational spaces, quantum computing approaches are emerging as a promising alternative.

The MOF-74 family has emerged as a standout candidate for post-combustion CO$_2$ capture due to its high density of open-metal sites—undercoordinated metal centers that serve as strong, reversible adsorption sites for CO$_2$ molecules. These open-metal sites dramatically enhance uptake capacity, allowing MOFs like Mg-MOF-74 to adsorb over 25\% of their weight in CO$_2$ under ambient conditions—far outperforming traditional sorbents such as zeolites \cite{yazaydin2009screening,mcdonald2012capture}. Within the MOF-74 family, substitution of the metal center (M = Mg, Mn, Fe, Co, Ni, Cu, Zn) provides a powerful handle for tuning adsorption performance \cite{lee2015small}. Despite structural similarity across M-MOF-74 variants, the specific open-metal site  significantly  impacts gas binding strength, selectivity, and kinetics. This makes it critical to understand, at the quantum level, how the choice of metal affects CO$_2$ interaction, a knowledge that is essential for rational material design.

\begin{figure}[!ht]
\centering
\includegraphics[width=0.8\columnwidth]{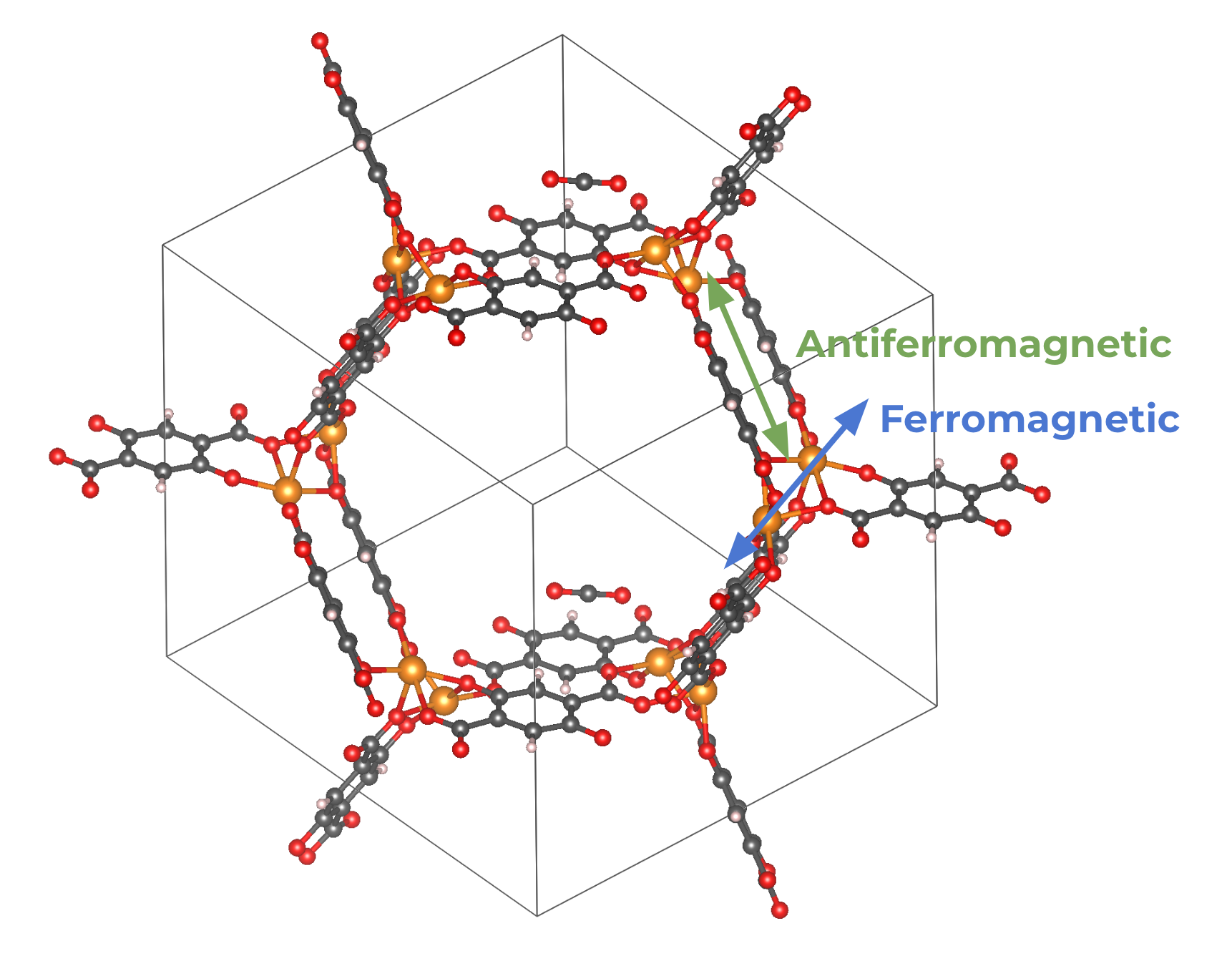}
\caption{The Fe-MOF-74 model considered in this work (H white, C black, O red, Fe orange).}
\label{fig:Fe-MOF}
\end{figure}

We consider Fe-MOF-74 as a prototypical system to study the CO$_2$ adsorption in MOFs using quantum computing. 
The structure of Fe-MOF-74 (Fig. \ref{fig:Fe-MOF}) is characterized by cylindrical cavities with exposed transition metal sites
where CO$_2$ can adsorb (two periodic repetitions of the adsorbed molecule are also displayed in the figure). 
This material exhibits quasi-linear metal oxide chains with ferromagnetic order; different chains are coupled antiferromagnetically, resulting in zero total magnetization in the cell.
This magnetic ordering has been reported both in computational \cite{canepa2013metal,maurice2013single} and experimental \cite{bloch2012hydrocarbon} studies. Similarly to other transition metal complexes in the MOF-74 family,
Fe-MOF-74 exhibits a Mott insulator behavior, where the localized d electrons experience strong on-site Coulomb and exchange correlations \cite{lee2015small}. This makes it an interesting 
test case for a quantum computing study.  Indeed, local or semi-local DFT functionals incorrectly predict this material to be nearly metallic 
instead of an insulator (considering a $\Gamma$-only sampling in the first Brillouin zone, the PBE bandgap is only 0.2 eV). An improved description of its electronic structure can be obtained by using the 
 empirical DFT+U approach \cite{anisimov1991band,anisimov1997first} or hybrid functionals \cite{krukau2006influence}. Quantum computers could offer a reliable description of this material without relying on empirical parameters.

Current quantum approaches have been typically restricted to molecular systems with a relatively small number of orbitals, limiting their application mainly to small molecules. Indeed, quantum studies of CO$_2$ adsorption in MOFs 
have so far relied on molecular fragment models of the MOF structure
 \cite{greene2022modelling}. Methodologies to study periodic materials are instead significantly less developed, although progress has been made for band structure calculations based on the tight-binding approximation \cite{cerasoli2020quantum}, optical properties of defects in solids \cite{ma2020quantum}, and other more general frameworks to compute ground and/or excited state properties of periodic solids \cite{liu2020simulating,yoshioka2022variational,ivanov2023quantum,clinton2024towards}.
Quantum applications to surface science problems have also been explored, typically employing reduced active spaces derived from relatively small localized basis sets \cite{gujarati2023quantum,di2024platinum,hualde2024quantum}. 
However, workflows based on the plane-wave basis set—the standard approach for most periodic electronic structure calculations—remain challenging and are generally considered targets for long term fault-tolerant quantum computing \cite{babbush2018low,su2021fault}.

In order to simulate the CO$_2$ adsorption in MOFs, in this work we develop a quantum workflow based on an active space selection starting from the plane-wave basis set. 
Following the methodology originally developed in Ref. \citenum{schafer2021local}
for classical simulations, the active space is built by selecting localized Wannier orbitals around the adsorption site (and on the molecule) and using natural orbitals
computed at the M\o{}ller-Plesset perturbation theory to second order (MP2) level
to converge the energy. It is shown that this approach achieves complete basis set-quality results for MP2 adsorption energies.  
The sizes of the active spaces built using this procedure are still too large for
quantum simulations and hardware experiments. To address this, we extract a smaller subspace that is used for numerical simulations based on the quantum number preserving (QNP) ansatz \cite{anselmetti2021local} within the variational quantum eigensolver (VQE) framework \cite{peruzzo2014variational}. This ansatz is also employed in quantum hardware experiments performed on IBM quantum devices. Since the largest active space considered in this work requires 28 qubits, the hardware implementation relies on the sample-based quantum diagonalization (QSD) approach \cite{robledo2024chemistry}, which enables substantially larger simulations than those accessible with the traditional VQE method. This study demonstrates a viable pathway toward simulating surface science problems using quantum computing, with the potential to extend this methodology to a broader range of challenging problems in materials science.


\section{Methods}

\subsection{Density functional theory calculations}

As a first step in this work, we performed density functional theory (DFT) calculations with two main objectives: (i) to validate the initial structural model, and (ii) to evaluate the predictive power of DFT by examining the variability of computed properties across different functionals, thereby motivating the adoption of quantum computing as a more robust predictive tool.
While the quantum study presented in this work focuses exclusively on Fe-MOF-74, the analysis based on density functional theory (DFT) also includes Mg-MOF-74. Although Mg-MOF-74 likely exhibits weaker electronic correlation due to the absence of 
d electrons, its inclusion is useful for validating our initial computational protocol and for gaining a clearer understanding of the range of predictions obtained from different DFT functionals.

We begin by considering the Fe-MOF-74 and Mg-MOF-74 structures reported in Ref. \citenum{lee2015small}. In that study, the initial primitive cell was derived from the experimentally known structure of Zn-MOF-74 \cite{rosi2005rod}, which was first optimized at the PBE level. Variants of MOF-74 were then generated by substituting Zn with other metal centers, including Fe and Mg. The initial adsorption geometries of CO$_2$ and other molecules were obtained using canonical Monte Carlo simulations with classical force fields. These geometries were subsequently optimized at the DFT level using the non-local vdW-DF2 functional \cite{lee2010higher}.

To begin, we reoptimized the structures of Fe-MOF-74 and Mg-MOF-74 from Ref. \citenum{lee2015small} using the VASP code \cite{kresse1994norm,kresse1996efficient}. For this purpose, we followed as closely as possible the computational protocol outlined in Ref. \citenum{lee2015small}. While that reference provides the main computational parameters, it does not include the full details of the calculations.

The atomic positions and cell parameters of the Fe-MOF-74 framework were reoptimized using the PBE+U functional with a Hubbard U value of 4 eV; the inclusion of the Hubbard correction was not required in the reoptimization of Mg-MOF-74. We employed a kinetic energy cutoff of 1000 eV, $\Gamma$-point sampling of the Brillouin zone, and a force convergence threshold of 0.01 eV/Å. For Fe-MOF-74 (Mg-MOF-74) the atomic positions changed by no more than 0.06 (0.07) \AA\, and the cell volume varied by just 0.01\% (0.9\%) compared to the original model in Ref. \citenum{lee2015small}. These minimal deviations confirm that our computational setup is consistent with previous work.

The geometry of the adsorbed CO$_2$ molecule was subsequently optimized using the non-local vdW-DF2 functional \cite{lee2010higher}, while keeping the MOF frameworks fixed. The atomic positions of CO$_2$ changed by no more than 0.07 (0.06) \AA\ for Fe-MOF-74 (Mg-MOF-74).
Although relaxing the framework might influence the calculated adsorption energy, our primary focus is on developing a quantum workflow for simulating \ce{CO2} adsorption in porous materials. Therefore, the approach adopted in Ref. \citenum{lee2015small} is sufficiently accurate for the purposes of this study.

The adsorption energy, which is the most relevant property for this study, is computed as: 
\begin{eqnarray}
-\Delta E = E_{\mathrm{CO_2@MOF}} - E_{\mathrm{MOF}} - E_{\mathrm{CO_2}}, \label{eq:Eads}
\end{eqnarray}
where $E_{\mathrm{CO_2@MOF}}$ denotes the ground-state total energy of the adsorbed system,
$E_{\mathrm{MOF}}$ that of the clean MOF, and $E_{\mathrm{CO_2}}$ the energy of the CO$_2$ molecule.
A truly predictive method for computing $\Delta E$ should be able to achieve chemical accuracy (1 kcal/mol$\approx$ 4.2 kJ/mol) in a systematic and reliable manner. The main challenges in applying DFT to adsorption in materials such as Fe-MOF-74 arise from its known limitations in describing van der Waals interactions and the strong electronic correlations associated with d orbitals, which have motivated the development of a range of empirical approaches \cite{klimevs2012perspective,anisimov1997first}.

In order to study the dependence of $\Delta E$ on the DFT approximation—and to motivate the adoption of higher-accuracy methods such as those based on quantum computing—we chose a set of representative functionals with
different characteristics but all including van der Waals (vdW) corrections at some level of approximation, as
required to describe molecular adsorption on surfaces.
Beyond the non-local van der Waals functional vdW-DF2 \cite{lee2010higher} we considered the GGA (generalized gradient approximation) PBE functional
\cite{perdew1996generalized} with Grimme's D3 vdW corrections \cite{grimme2010consistent}, the meta-GGA SCAN functional
with a non-local rVV10 term to keep into account the vdW contribution \cite{vydrov2010nonlocal, peng2016versatile}, the
HSE06 \cite{krukau2006influence} and B3LYP \cite{stephens1994ab} hybrid functionals with D3 corrections. For the HSE06 calculation
we additionally considered the Becke-Johnson (BJ) damping function \cite{grimme2011effect}. For the hybrid functionals the
parameters for the damping have to be provided explicitly in the Vasp input, and have been obtained 
from Grimme's group web page \cite{dampparam}. All of these functionals have been benchmarked using the same optimized geometry.
 
As also discussed in the next section, the use of DFT+U is necessary to prevent Fe-MOF-74 from becoming 
nearly metallic at the vdW-DF2 and PBE level and from significantly underestimating the gap at the SCAN level \cite{anisimov1991band,anisimov1997first}. This is an empirical
approach that requires the choice of the U parameter, often selected to improve the description of certain properties at the expense of others. 
For the vdW-DF2, PBE and SCAN+rVV10 functionals we included an Hubbard U correction with a value of 4 eV (the Hubbard correction is applied exclusively to the d orbitals of Fe.).
This value was proposed in Ref. \citenum{wang2006oxidation} to reproduce the experimental oxidation energy of transition metal oxides (namely this value was not obtained specifically for Fe-MOF-74).
An alternative value of U=6.5 eV was specifically obtained for Fe-MOF-74 in Ref. \citenum{mann2016first} using the linear response
approach of Cococcioni and de Gironcoli \cite{cococcioni2005linear}. According to the results of Ref. \citenum{mann2016first}
the adsorption energy of CO$_2$ in Fe-MOF-74 with U=6.5 eV does not change by more than 1-2 kJ/mol compared to U=4 eV.

\subsection{Active space generation}\label{sec:ASmethod}

Periodic calculations based on plane-waves and pseudopotentials involve a very large basis set. For the DFT benchmarks performed here, this corresponds to more than 100,000 plane-waves for semi-local/meta-GGA functionals (1000 eV cutoff) and almost 27,000 plane-waves for hybrid functionals, where the default cutoff (400 eV) was used because of the higher computational cost. Quantum simulations involving 10$^4$-10$^5$
basis elements are clearly impractical for near and medium term quantum computing applications. Therefore, compression into a smaller active space will be essential for enabling practical applications.

Active space techniques are relatively well established for molecular simulations \cite{sayfutyarova2017automated} but less developed for materials applications. New methodological advancements based on periodic boundary conditions could significantly enhance the potential of quantum computing in materials science. 
Here we use the approach developed in Ref. \citenum{schafer2021local}. While all the details of this methodology are described
in detail in this original reference,
we summarize here the main steps of the computational procedure:
\begin{enumerate}
    \item Initial Hartree-Fock ground-state calculation.
    \item Transforming the occupied states into Wannier functions in the form of Intrinsic Bond Orbitals \cite{schafer2021surface, wockinger2024exploring}, which involves applying a unitary transformation to the occupied Bloch orbitals to obtain a localized orbital representation.    
    \item Selection of the Wannier orbitals localized around the site of interest (the adsorption site around a Fe atom in this case). These orbitals are then recanonicalized.
    \item Calculation of the MP2 natural orbitals for the unoccupied state manifold; only the selected occupied orbitals are included in the reduced one-electron density matrix calculation. It was demonstrated that the use of natural orbitals provides a much faster convergence of the correlation energy at the MP2 and coupled-cluster levels compared to the simple use of Hartree-Fock orbitals \cite{grüneis2011natural}. The selected subspace is finally recanonicalized and used to define one and two-electron integrals.
\end{enumerate}

The original active space approach presented in Ref. \citenum{schafer2021local} was developed with the primary goal of embedding coupled cluster theory within the random phase approximation (RPA). However, the underlying idea is general and can be applied to embed any high-level method $\mathrm{M}_1$ (including high-accuracy quantum computing approaches) within a lower-level theory $\mathrm{M}_2$. The embedded correlation energy in the unit cell is approximated as
\begin{eqnarray}
 E^{\mathrm{M}_1:\mathrm{M}_2 \mathrm{corr}} &=& E^{\mathrm{M}_1 \mathrm{corr}}_{\mathrm{AS}} + E^{\mathrm{M}_2 \mathrm{corr}}_{\mathrm{R}} + E^{\mathrm{M}_2 \mathrm{corr}}_{\mathrm{AS}\leftrightarrow \mathrm{R}} \nonumber \\
&=& E^{\mathrm{M}_1 \mathrm{corr}}_{\mathrm{AS}} + E^{\mathrm{M}_2 \mathrm{corr}} - E^{\mathrm{M}_2 \mathrm{corr}}_{\mathrm{AS}},
\end{eqnarray}
where $E^{\mathrm{M}_1 \mathrm{corr}}_{\mathrm{AS}}$ denotes the M$_1$ correlation energy within the active space (AS), $E^{\mathrm{M}_2 \mathrm{corr}}_{\mathrm{R}}$ corrsponds to the M$_2$ correlation energy in the rest of the cell (R), and $E^{\mathrm{M}_2 \mathrm{corr}}_{\mathrm{AS}\leftrightarrow \mathrm{R}}$ captures the interaction energy between AS and R, assumed to be well approximated at the lower level of theory  M$_2$.
Under this assumption the second line of the equation follows from the identity
$E^{\mathrm{M}_2 \mathrm{corr}}_{\mathrm{R}} + E^{\mathrm{M}_2 \mathrm{corr}}_{\mathrm{AS}\leftrightarrow \mathrm{R}}=E^{\mathrm{M}_2 \mathrm{corr}} - E^{\mathrm{M}_2 \mathrm{corr}}_{\mathrm{AS}}$.
For simplicity, in this work, the active space—comprising the d electrons and treated using MP2 and quantum computing (QC) techniques—is embedded within Hartree–Fock (HF) theory. While more advanced embeddings (e.g. based on the RPA) could potentially offer improved accuracy, HF is sufficient for this proof-of-concept study. Moreover, extending the approach beyond HF is straightforward. Since HF yields zero correlation energy for the lower-level theory $\mathrm{M}2$, the adsorption energy (Eq. \ref{eq:Eads}) can be expressed as
\begin{eqnarray}
    \Delta E = \Delta E^{\mathrm{HF}} + \Delta E^{\mathrm{M}_1\mathrm{corr}}_{\mathrm{AS}}, \label{eq:Eadstot}
\end{eqnarray}
where $\Delta E^{\mathrm{HF}}$ is the Hartree-Fock adsorption energy and
\begin{eqnarray}
-\Delta E^{\mathrm{M}_1 \mathrm{corr}}_{\mathrm{AS}} = E_{\mathrm{CO_2@MOF},\mathrm{AS}}^{\mathrm{M}_1\mathrm{corr}} - E_{\mathrm{MOF},\mathrm{AS}}^{\mathrm{\mathrm{M}_1corr}} - E_{\mathrm{CO_2},\mathrm{AS}}^{\mathrm{M}_1\mathrm{corr}} \label{eq:Eadscorr}
\end{eqnarray}
is the correlation contribution to the adsorption energy, evaluated in the active space. Embedding in a higher-level theory such as RPA would require corresponding calculations for both the active space and the full system, which would be computationally tractable for the case considered here.

The active spaces are constructed to include orbitals localized on the iron atom at the adsorption site, along with those of nearby atoms, depending on the size of the problem that can be feasibly treated. The selected iron atom is in a high-spin (quintet) state and couples antiferromagnetically to some of the other iron centers in the framework. A straightforward spin-polarized active space construction centered on a single Fe atom would result in both the active and frozen spaces being in high-spin states. This leads to several complications. In particular, the interaction between these two high-spin subsystems likely involves substantial static correlation, creating a conceptual inconsistency, since the active space is intended to capture the strong correlation effects in the system. Capturing the full magnetic coupling among the iron atoms would, in principle, require inclusion of orbitals from all Fe centers and possibly the neighboring ligands that connect them. However, such an approach would be computationally demanding and, in many cases, unnecessary, as the adsorption process is predominantly localized at a single metal site. For this reason, we generated the active space using a spin-restricted calculation. It is important to emphasize that this spin-restricted treatment is used solely for the purpose of generating the active space. Within the active space itself, the system is still allowed to be spin-polarized. That is, subsequent quantum computations—or classical full configuration interaction calculations—performed within the active space can recover the correct local high-spin state. Although this approach does not fully describe the global magnetic ordering, such correlations can be progressively restored by systematically enlarging the active space to include orbitals from additional metal centers and nearby atoms.

Since the active spaces generated by this procedure are still relatively large for the capabilities of current quantum computers, we define a reduced working space by selecting a subset of orbitals. Although this is not a physically rigorous subspace, it serves as a practical testbed for quantum simulations and hardware experiments.

\subsection{Quantum simulations based on the quantum number preserving ansatz}

\begin{figure}[!ht]
\centering
\includegraphics[width=0.37\columnwidth]{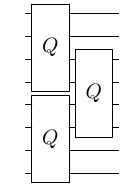}
\caption{The QNP fabric circuit proposed in Ref. \citenum{anselmetti2021local}.}
\label{fig:QNP}
\end{figure}

Once active spaces of practical size were identified, quantum simulations on classical hardware were performed using the Variational Quantum Eigensolver (VQE).
The VQE is a hybrid algorithm that combines quantum and classical computations and is widely considered one of the most promising approaches for practical applications in the noisy intermediate-scale quantum (NISQ) era \cite{peruzzo2014variational,tilly2022variational}.
Implementing VQE in practice requires selecting an ansatz that balances accuracy in representing the target ground-state with manageable circuit depth, ideally scaling polynomially. Several ansatz choices have been explored in the literature \cite{tilly2022variational}. For simulations
in chemistry and materials science, it is particularly advantageous to use ansatzes that enforce particle number and spin symmetries. 
This is the case for the quantum number preserving (QNP) ansatz proposed in Ref. 
\citenum{anselmetti2021local}.
This approach employs a two-parameter, four-qubit gate $Q(\phi,\theta)$, constructed from two components: The spin-adapted spatial orbital rotation gate QNP$_\mathrm{OR}$($\phi$) and the diagonal pair exchange gate QNP$_\mathrm{PX}$($\theta$). Implementing QNP$_\mathrm{OR}$($\phi$) requires 4 CNOT gates, while QNP$_\mathrm{PX}$($\theta$) uses 13 CNOTs. An example of an 8-qubit QNP layer is illustrated in Fig. \ref{fig:QNP}. Notably, the QNP ansatz has demonstrated strong expressiveness at relatively shallow depths and its parameter optimization is less susceptible to barren plateaus \citenum{anselmetti2021local,mcclean2018barren}.

To evaluate the accuracy of the QNP ansatz, simulations were performed on a noiseless quantum simulator, Quicksilver by QC Ware, that runs on classical (GPU) hardware. This Fermionic simulator uses a double-factorized (DF) representation of the Hamiltonian, as described in Ref. \citenum{cohn2021quantum}.
Within this formalism, a rank-truncated decomposition of the two-electron integrals is employed to reduce the number of terms in the Hamiltonian.
The truncation is controlled by the integer parameter $N_{\mathrm{DF}}$, which can be chosen to balance accuracy and computational efficiency; its maximum value corresponds to the square of the number of orbitals.

For the smallest active space considered here, corresponding to the CO$_2$ molecule, the optimization of the parameters in the QNP ansatz is straightforward and was directly obtained using the conjugate gradient algorithm. 
The optimization of the QNP ansatz for the active spaces selected for MOF and CO2@MOF is significantly more complex because of the relatively large numbers of qubits (20 and 28, respectively) and the presence of several local minima in the energy landscape.
To overcome these challenges we have implemented a basin-hopping procedure\cite{wales1997global} that loops through the following steps:
\begin{enumerate}
    \item Starting from an initial configuration (namely, a set of QNP parameters), the minimum of the current basin is obtained using the conjugate gradient (CG) algorithm. To speed up calculations, a loose convergence threshold was used for the CG algorithm ($10^{-4}$ on the gradient for MOF and $10^{-3}$ for CO$_2$@MOF) and underconverged values for $N_{\mathrm{DF}}$ in the double-factorized Hamiltonian (41 for MOF and 28 for CO$_2$@MOF).
    \item A relatively large perturbation is applied to the optimized configuration in order to move out of the current basin.
    \item A new optimization with loose convergence parameters is performed to obtain the minimum value of the energy in the current basin.
    \item If the new energy is lower than in the previous basin, the new configuration is automatically accepted. If the new energy is higher, a Metropolis-type criterion is applied to decide whether the new configuration is accepted or not.
    \item The procedure is repeated from step 2.
    \item Once the procedure is completed, the lowest minimum found is reoptimized using tight convergence parameters for the CG algorithms and for the Hamiltonian representation. 
\end{enumerate}

As discussed below, this procedure makes it possible to achieve chemically accurate results even for the MOF and CO$_2$@MOF active spaces.

\subsection{Quantum hardware experiments using sample-based quantum diagonalization}\label{sec:SQDmethod}

Quantum hardware experiments based on a straightforward application of the Variational Quantum Eigensolver face several practical challenges on today’s noisy devices. Even assuming that the variational parameters have been pre-optimized, the number of measurements required to evaluate the energy is impractically high — and increases further when noise mitigation is considered. As a result, practical applications have been so far limited to at most 12 qubits using simple ansatzes \cite{google2020hartree,zhao2023orbital}.
Accordingly, directly performing VQE-based hardware experiments of the selected active spaces of MOF (20 qubits) and CO$_2$@MOF (28 qubits) is currently out of reach. To overcome these difficulties, we apply the QNP ansatz within the framework of 
sample-based quantum diagonalization (SQD), a hybrid quantum-classical algorithm 
recently introduced by the IBM Quantum research team. SQD has already been successfully employed in chemistry simulations with up to about 80 qubits \cite{robledo2024chemistry,kaliakin2024accurate,shajan2024towards,yu2025quantum}.
This approach is based on earlier work on quantum-selected configurations interaction \cite{kanno2023quantum,nakagawa2024adapt}; other analogous variants of SQD are also available on the IBM Quantum Platform \cite{pellow2025hivqe,shirai2025enhancing}.

Sample-based quantum diagonalization and other quantum-enhanced selected CI approaches begins by preparing an approximate ground-state wavefunction on a quantum processor, using methods such as the variational quantum eigensolver (VQE) \cite{robledo2024chemistry} (specifically, in this work the QNP ansatz will be used) or quantum Krylov \cite{yu2025quantum}. 
This state $\Psi$ can be seen as a linear combination of computational states
\begin{eqnarray}\label{SQDgs}
    |\Psi \rangle = \sum_{\mathbf{x}} a_{\mathbf{x}} |\mathbf{x}\rangle, 
\end{eqnarray}
where $\mathbf{x}\in \{0,1\}^{N_{\mathrm{q}}}$ is a $N_{\mathrm{q}}$-bit string. Within the Jordan-Wigner
mapping these bit strings represent electronic configurations where 0 and 1 correspond to unoccupied and occupied orbitals, respectively. By measuring the state in the computational basis, each bit string can be sampled with
probability $|a_{\mathbf{x}}|^2$. Sampling these bit strings in a quantum experiment helps to identify a subset of electronic configurations that play a dominant role in describing the ground-state. The Hamiltonian can then be diagonalized classically in this subspace, whose size is significantly reduced with respect to the full Hilbert space. SQD provides the following advantages: 
\begin{itemize}
    \item Noise resilience: Because quantum hardware is used only for identifying the subspace and not for the energy evaluation, the method is inherently resilient to both noise and sampling error. Additionally, the energy is variational even in the presence of quantum noise.
    \item Lower depth/complexity of the quantum circuits: Ideally $\Psi$ in Eq. \ref{SQDgs} should correspond to the exact ground-state; in practice, it is sufficient for $\Psi$ to have a support similar to that of the true ground-state. Indeed, as long as the relevant subspace of electronic configurations is adequately sampled, an accurate estimate of the energy can be obtained. This typically requires less deep and complex quantum circuits for state preparation. 
    \item Takes advantage of sparsity: If the quantum state is sparse (namely, a relatively small number of bit strings $\mathbf{x}$ contribute to $\Psi$), then the SQD approach becomes particularly efficient. However, if this is initial state is not sufficiently sparse, the computational load on the classical processor can quickly become impractical.
\end{itemize}

While the classical diagonalization guarantees noise resilience and variationality of the ground state energy, the sampling of bit strings
from noisy hardware provide several electronic configurations that do not conserve the total or spin-resolved number of electrons.
Although these unphysical configurations could be discarded, doing so would result in the loss of a substantial amount of data produced in the quantum experiment that may still contain useful information.
To overcome this issue, here we use the self-consistent recovery procedure
proposed in Ref. \citenum{robledo2024chemistry}.
In each iteration of this procedure, the algorithm scans the configuration set $\{\mathbf{x}\}$ and identifies those states with the correct number of electrons $N_{\mathbf{x}} = N_{\mathrm{e}}$ (where $N_{\mathbf{x}}$ is the Hamming weight of $\mathbf{x}$ and $N_{\mathrm{e}}$ the number of electrons). For configurations with $N_{\mathbf{x}} > N_{\mathrm{e}}$ or $N_{\mathbf{x}} < N_{\mathrm{e}}$, a correction is applied by flipping \( |N_{\mathbf{x}} - N_{\mathrm{e}}| \) bits. These bits are selected from the set of occupied or unoccupied spin-orbitals (respectively), guided by a probability distribution that increases monotonically with $|x_{p\sigma} - n_{p\sigma}|$, where  $n_{p\sigma}$ is the average occupation of orbital \( p\sigma \) from the previous round. This process yields a refined set of configurations and may be repeated iteratively to achieve convergence.

\section{Results and discussion}

\subsection{Density functional theory benchmark}

In Table \ref{tab:adsE} we present the results for the adsorption energies 
computed considering different DFT functionals. Experimental results from Ref. \citenum{queen2014comprehensive} are also provided in the last line. The quantity measured experimentally is the isosteric heat of adsorption that in the low-coverage limit compares to the computed enthalpy of adsorption rather than the energy of adsorption in Eq. \ref{eq:Eads}.
For a straightforward comparison, the energies $E$ computed at the DFT level should be corrected for quantum nuclear zero-point energy (ZPE) and thermal energy (TE) to obtain the enthalpy $H$. 
Analogously to Eq. \ref{eq:Eads} the enthalpy of adsorption is defined as 
$-\Delta H(T) = H(T)_{\mathrm{CO_2@MOF}} - H(T)_{\mathrm{MOF}} - H(T)_{\mathrm{CO_2}}$. 
Rather than applying the corrections to each DFT result, in Table \ref{tab:adsE} we remove these contributions from the experimental value of the low coverage heat of adsorption using the corrections computed at the DFT level in Ref. \citenum{lee2015small}; this approach provides an experimental estimate of $\Delta E$. To be consistent with experiment the TE corrections were evaluated at room temperature (297 K).
The corresponding values are ZPE=1.7 kJ/mol and TE=1.9 kJ/mol for CO$_2$ in Fe-MOF-74, and  ZPE=2.2 kJ/mol and TE=1.6 kJ/mol for CO$_2$ in Mg-MOF-74. The experimental value is provided
as reference but it should be considered that both experimental uncertainties and approximate ZPE/TE corrections could influence it. 

All the functionals correctly predict stronger adsorption of \ce{CO2} in Mg-MOF-74 than Fe-MOF-74, as also highlighted by the last column of Table \ref{tab:adsE}; however, the relative stability is not always quantitatively consistent, as the relative energies
range from 4.7 to 10.1 kJ/mol.
Overall all the different DFT functionals considered here perform fairly well but with an evident variability among the different functionals and in several cases the deviations of the
computed adsorption energies with respect to experiment are well beyond chemical accuracy. In particular, the largest deviations are obtained for HSE06-D3 with Becke-Johnson damping. However, highly accurate adsorption energies are required for reliable predictions of kinetic parameters. This motivates the use
of quantum computing approaches to enhance accuracy and achieve a comparison with experiment not biased by the specific 
choice of the functional.

\begin{table*}[tb!]
\centering
\resizebox{\textwidth}{!}{
\begin{tabular}{|c|c|c|c|c|c|}
\hline  \hline
Functional       & Hubbard U   &  Type & CO$_2$ in Mg-MOF-74  & CO$_{2}$ in Fe-MOF-74 & relative  \\ 
 & on Fe (eV) & & $\Delta E$ (kJ/mol) & $\Delta E$ (kJ/mol) & stability (kJ/mol)  \\ \hline \hline
vdW-DF2 & 4 & Non-local vdW   & 45.7 (-1.6)  & 37.9 (1.1) & 7.8 \\
vdW-DF2 (Ref. \cite{lee2015small}) & 4 & Non-local vdW   & 45 (-2.3) & 38 (1.2) & 7 \\
PBE-D3 & 4 & GGA+vdW correction  & 40.7 (-6.6) & 36.0 (-0.8) & 4.7 \\  
SCAN+rVV10 &  4 & metaGGA+non-local vdW   & 51.0 (3.7) & 43.5 (6.7) & 7.5 \\
HSE06-D3 & / & Range-separated hybrid+vdW corr. & 40.5 (-6.8) & 31.0 (-5.8) & 9.5  \\
HSE06-D3 (BJ) & / & Range-separated hybrid+vdW corr. & 60.0 (12.7) & 50.8 (14.0) & 9.2 \\
B3LYP-D3 & / & Unscreened hybrid+vdW corr. & 45.9 (-1.4) & 35.8 (-1.0) & 10.1  \\
\hline \hline
Exp. (Ref. \citenum{bloch2012hydrocarbon}) & / & Adsorption isotherms  & 47.3 & 36.8 & 10.5 \\
\hline \hline
\end{tabular}}
\caption{Total adsorption energies $\Delta E$ (Eq. \ref{eq:Eads}) of CO$_2$ in Mg-MOF-74 and Fe-MOF-74 as computed with different functionals. The values in parentheses correspond to the deviation of the computed value with respect to experiment. BJ stands
for Becke-Johnson (BJ) damping \cite{grimme2011effect}. The last column shows the difference $\Delta E$(CO$_2$ in Mg-MOF-74)
- $\Delta E$(CO$_2$ in Fe-MOF-74). As described in the main text, the zero-point energy and thermal energy contributions computed at DFT level \cite{lee2015small} have been removed from the heats of adsorption to obtain the experimental estimates of $\Delta E$ in the last line.}
\label{tab:adsE}
\end{table*}

\subsection{Construction of active spaces}

Following the procedure described in Sec. \ref{sec:ASmethod}, the
actives space for CO$_2$@MOF is selected to include all the occupied orbitals of CO$_2$, and all the localized occupied orbitals associated to the Fe atom in the adsorption site and the four surrounding oxygen atoms. 
A consistent approach is then used for the CO$_2$ and MOF sepearately. 
Since the orbitals are recanonicalized, it should be noticed that the localization procedure on CO$_2$ does not have any effect for this system.
A systematic convergence test with respect to the number of occupied orbitals is generally required to ensure the accuracy of the embedded active space, as demonstrated in Ref. \citenum{schafer2021local}. However, for the purposes of this work—focused on demonstrating a quantum simulation workflow for carbon capture in MOFs—the current choice provides a sufficiently accurate and practical starting point.

Convergence with respect to the size of the unoccupied-state manifold is examined in greater detail, as this has important implications in determining whether converged adsorption energies can be reliably obtained. By initially considering a kinetic energy cutoff of 320 eV,
the sizes of the different active spaces and the corresponding MP2 correlation only adsorption energies (Eq. \ref{eq:Eadscorr}) are shown in 
Table \ref{tab:adsEactive}. To select the size of the full active space 
different thresholds were used for the natural orbital (NO) occupations.
Applying the same threshold across the three systems (CO$_2$, MOF, and CO$_2$@MOF) ensures reasonable consistency among the active spaces and error cancellation when computing the adsorption energy. 
While a threshold of $10^{-3}$ is not sufficient to achieve convergence, all other thresholds yield consistent results (it is important to notice that a difference of 1 kJ/mol represents a small energy scale, well within the range of chemical accuracy, typically defined as 1 kcal/mol$\approx$4.2 kJ/mol). 
Since MP2 is a correlated method, these findings suggests that converging quantum computing–quality adsorption energies may be achievable, although definitive validation will only be possible with the advent of large-scale quantum computers.

By summing the HF and MP2 contributions to the adsorption energy (see Eq. \ref{eq:Eadstot}), 
a value $\Delta E = 28$ kJ/mol is obtained. This is unsatisfactory
when compared to the experimental value of 36.8 kJ/mol but several factors should be taken into account: (1) MP2 is an approximate method that does not systematically achieve chemical accuracy; (2) in this case MP2 is simply embedded in HF rather than in  
 another correlated approach, such as RPA; (3) for a fully quantitative prediction, it might be necessary to include orbitals from additional atoms; (4) the MP2 calculation is spin restricted, as the main purpose of this approach was to build the active space. 

The size of the active spaces in Table \ref{tab:adsEactive} are still relatively large for the capabilities of current quantum computers. However, a significant reduction in dimensionality is achieved compared 
to the total number of orbitals involved in the plane-wave
calculation. Indeed, considering the total number of valence electrons and the size of the basis set, the full active space for CO$_2$@MOF would be  
(322, 18237). The quantum simulation of such a large active space could be challenging even in the long term. In this context, the reduction of the number of orbitals by several orders of magnitude that can be achieved with our procedure will be instrumental in future applications
of quantum computing in materials science.  

\begin{figure}[!ht]
\centering
\includegraphics[width=\columnwidth]{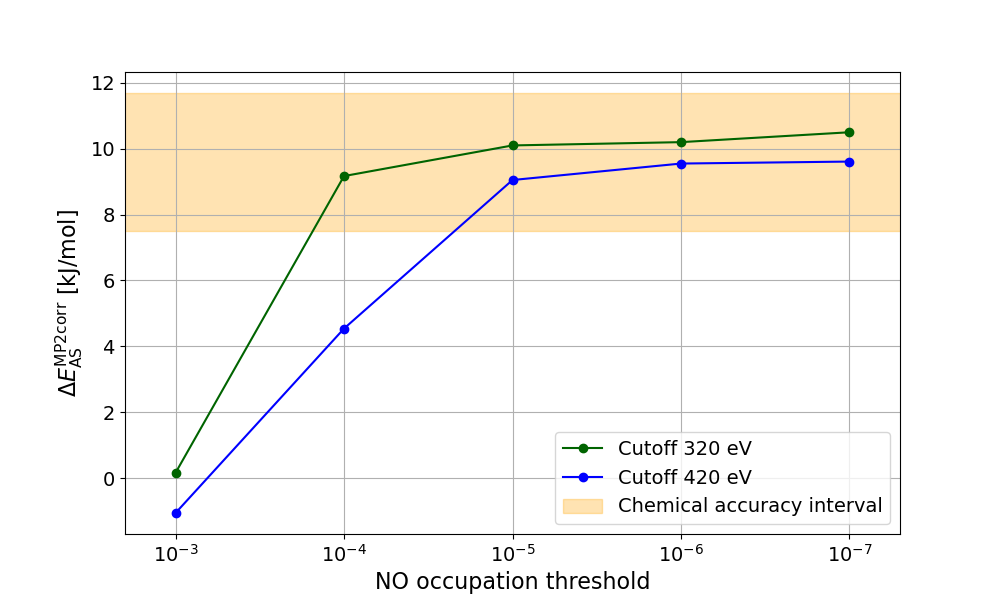}
\caption{Convergence of the contribution of the MP2 correlation energy to the adsorption energy in the active space $\Delta E^{\mathrm{MP2}\mathrm{corr}}_{\mathrm{AS}}$ (see Eq. \ref{eq:Eadscorr}) as a function of the threshold used to truncate MP2 natural orbitals. Results are displayed for two different kinetic energy cutoffs used in the underlying Hartree-Fock calculation. The shaded area denotes the chemical accuracy interval around
the most converged point obtained for a $10^{-7}$ threshold and 420 eV cutoff.}
\label{fig:cutoffconv}
\end{figure}

Compared to standard quantum chemical calculations based on localized basis sets,
the orbitals in the active spaces in Table \ref{tab:adsEactive} have been built  
from an underlying ``overcomplete'' plane-wave basis set. This implies that the interaction energy results
could already achieve complete basis set accuracy without relying on extrapolation techniques. To quantitatively support this point, we generated additional active spaces using an increased plane-wave cutoff of 420 eV. This higher cutoff has only a modest effect on the Hartree–Fock (HF) adsorption energy, which shifts from 18.0 kJ/mol (at 320 eV) to 18.7 kJ/mol. While such a change may appear negligible, it is important to note that the convergence of correlated methods—such as MP2 or quantum computing-based approaches—is not usually guaranteed by HF-level convergence. Nevertheless, as shown in Fig. \ref{fig:cutoffconv}, the convergence of the adsorption energy with respect to the underlying plane-wave basis set remains well within chemical accuracy (the detailed numerical results using the 420 eV cutoff are provided in Table S1 in the Supplementary Information). This finding is significant, as it demonstrates that a plane-wave-based approach can yield rapid and reliable basis set convergence, particularly in cases where favorable error cancellation occurs in energy differences.

The generated active spaces account for both dynamical correlation and any static correlation that may arise from the d electrons (d-orbitals are included in this subspace, see discussion in Sec. \ref{sec:reduced}). However, even the active spaces in Table \ref{tab:adsEactive} are still too large
to be simulated on current quantum hardware. For this reason, in the next section (Sec. \ref{sec:reduced}) we extract significantly smaller active spaces to proceed with the project.

\begin{table*}[tb!]
\centering
\resizebox{0.7\textwidth}{!}{
\begin{tabular}{|c|l|l|l|l|}
\hline  \hline
NO occupation   & active space   & active space & active space  &  $\Delta E$  \\ 
threshold & CO$_2$    & MOF & CO$_2$@MOF & kJ/mol  \\ 
  \hline \hline
10$^{-3}$ & (16,28) & (46,72) & (62,100)  & 0.160 \\
10$^{-4}$ & (16,62) & (46,151) & (62,214) & 9.17 \\
10$^{-5}$ & (16,116) & (46,300) & (62,418) & 10.1 \\
10$^{-6}$ & (16,200) & (46,524) & (62,724) & 10.2 \\
10$^{-7}$ & (16,308) & (46,854) & (62,1170) & 10.5 \\ 
\hline \hline
\end{tabular}}
\caption{Contribution of the MP2 correlation energy in the active space to the adsorption energy $\Delta E^{\mathrm{MP2} \mathrm{corr}}_{\mathrm{AS}}$ of CO$_2$ in Fe-MOF-74 as obtained with active spaces of different sizes. These results were obtained with a kinetic energy cutoff of 320 eV.}
\label{tab:adsEactive}
\end{table*}

\subsection{Reduced active space for quantum simulations and hardware experiments}\label{sec:reduced}

In order to build a smaller working active space, a subset of orbitals is extracted 
from the subspace corresponding to the $10^{-4}$ threshold in the NO occupancies (Table \ref{tab:adsEactive}). This reduced active space is mainly meant for the practical purpose
of performing quantum simulations and hardware experiments, and to include d orbital components, that
might involve some level of strong correlation. However, this active space is not expected to
be quantitatively accurate. 

\begin{figure}[!ht]
\centering
\includegraphics[width=0.7\columnwidth]{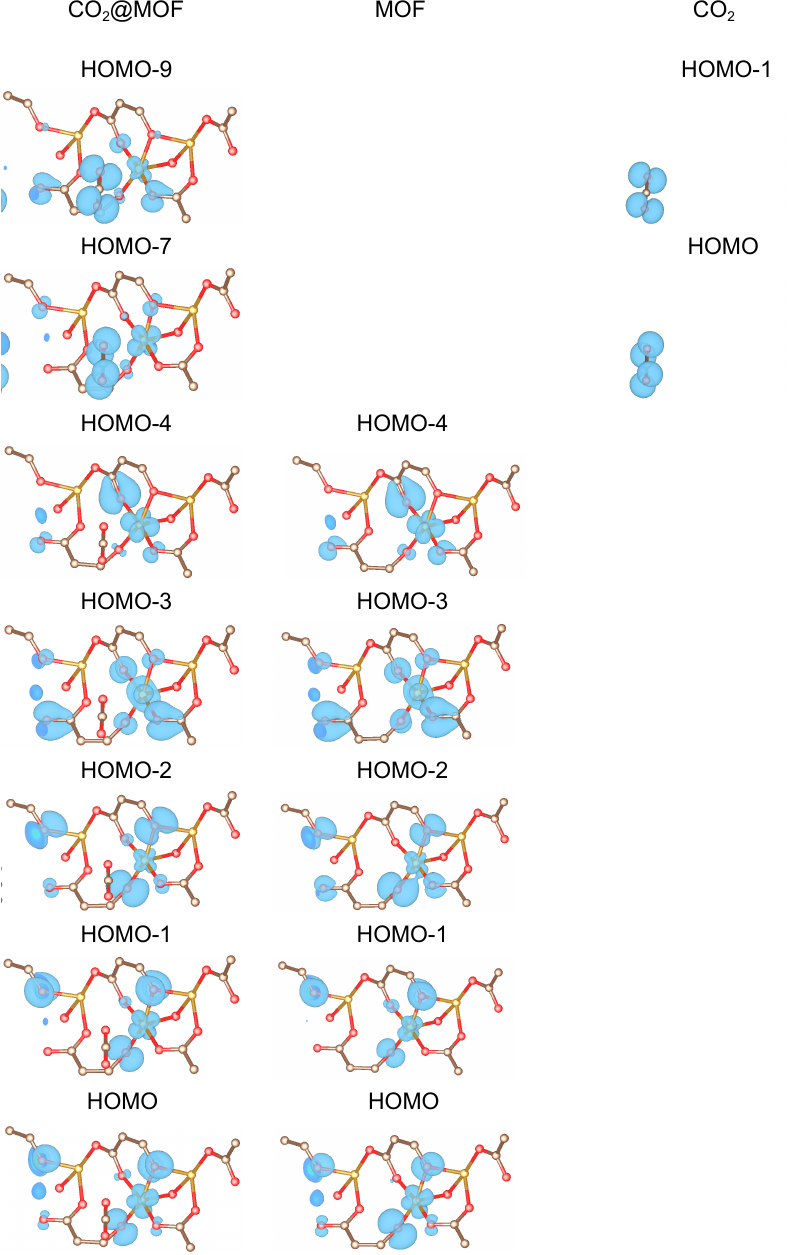}
\caption{Partial charges of the occupied orbitals included in the reduced active space of CO$_2$@MOF, MOF, and CO$_2$. The MOF and CO$_2$ orbitals are depicted on the side of the corresponding orbital in the adsorbed system. To simplify the visualization, only some of the atoms surrounding the adsorption site are displayed, but numerical calculations have been performed on the full periodic system.}
\label{fig:MOFCO2occ}
\end{figure}

\begin{figure}[!ht]
\centering
\includegraphics[width=0.7\columnwidth]{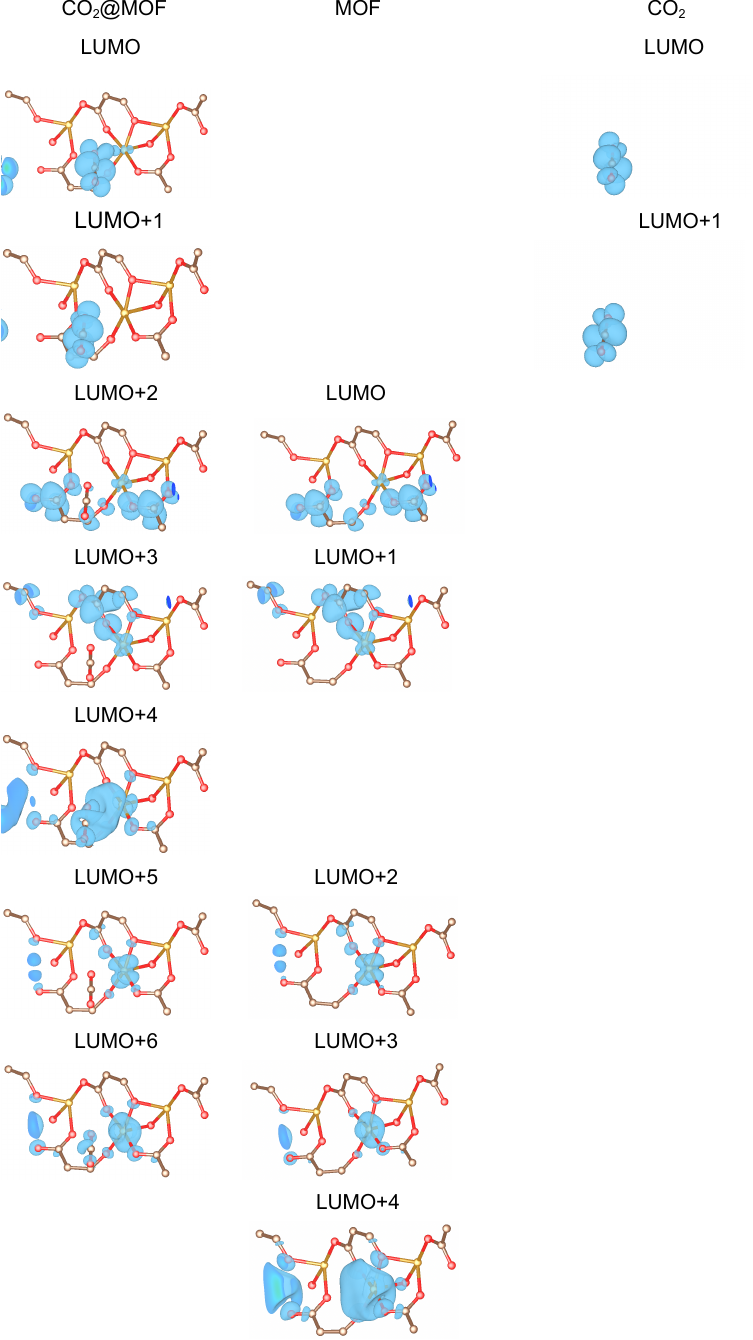}
\caption{Partial charges of the unoccupied orbitals included in the reduced active space of CO$_2$@MOF, MOF, and CO$_2$. The MOF and CO$_2$ orbitals are depicted on the side of the corresponding orbital in the adsorbed system. To simplify the visualization, only some of the atoms surrounding the adsorption site are displayed, but numerical calculations have been performed on the full periodic system.}
\label{fig:MOFCO2empty}
\end{figure}

Although the simulations are carried out under periodic boundary conditions, the Brillouin zone is sampled only at the $\Gamma$ point and the resulting orbitals are spatially localized. Therefore, in the discussion below, we adopt a molecular electronic structure terminology: the highest occupied state is referred to as the highest occupied molecular orbital (HOMO), with the next lower energy states labeled HOMO–1, HOMO–2, and so on. Similarly, the lowest unoccupied state is denoted as the lowest unoccupied molecular orbital (LUMO), followed by LUMO+1, LUMO+2, etc., for orbitals of increasing energy.
For the reduced active spaces of MOF and CO$_2$, only the orbitals around the band gap 
were selected, resulting in active spaces of (10,10) size for MOF and (4,4) for CO$_2$. 
The selected occupied and empty orbitals are depicted in Fig. \ref{fig:MOFCO2occ} and Fig. \ref{fig:MOFCO2empty}, respectively. 
Several of the active space orbitals of MOF clearly show some d orbital character for both the
occupied and empty states. The active space of CO$_2$ includes two degenerate occupied orbitals corresponding to the oxygen lone pairs and two degenerate non-bonding empty orbitals.

To be consistent with the two subsystems, the active space for the adsorbed system is chosen
to have a (14,14) size. The selected orbitals are depicted in Figs. \ref{fig:MOFCO2occ} and \ref{fig:MOFCO2empty} for occupied and empty states, respectively.
The five highest occupied orbitals of CO$_2$@MOF have an almost perfect one to one correspondence with the five highest occupied orbitals of the clean MOF. The CO$_2$ occupied orbitals for the standalone molecule can be
identified at HOMO-7 and HOMO-9 in the adsorbed system and present also some charge density localized on the substrate. For the conduction state subspace, the 7 lowest lying 
states were selected (as those are constructed to optimally converge the correlation energy, at least at the MP2 level). Each of those orbitals have an almost perfect counterpart in the MOF
or CO$_2$, but for LUMO+4. The partial charge corresponding to this orbital 
is delocalized between the molecule and the substrate, and provides an important contribution to the adsorption energy. 
As a result, no counterpart to this orbital is found in the ``monomers'', although the LUMO+4 of MOF is significantly more delocalized than the other selected orbitals.

Performing spin-restricted full CI on the reduced active spaces discussed in this section 
leads to a total (including the HF contribution as done in Eq. \ref{eq:Eadscorr}) adsorption energy of 43.4 kJ/mol. While this value is in fair agreement with the experimental value of 36.8 kJ/mol, such small active space sizes should not be considered reliable for systematic $\Delta E$ predictions. 

\subsection{Quantum simulations}\label{sec:QSresults}

Before testing the performance of the QNP ansatz in VQE simulations, it is important to select the specific spin state to simulate in the active spaces. As previously discussed, the Fe atom at the adsorption site in the periodic material is in a quintet state. The active spaces have been generated from spin restricted calculations but different spin multiplicities can be considered afterward. To establish the relative stability of the  
 different spin states we performed full CI calculations with the Pyscf code \cite{sun2018pyscf}. The reduced active spaces were considered, with sizes (4,4) for CO$_2$, (10,10) for MOF, and (14,14) for CO$_2$@MOF. 
The results in Table S2 of the Supplementary Information clearly show that the singlet states are the most stable with the triplet states providing slightly higher energies. The quintet state energies are significantly higher in energy and provide a negative adsorption energy $\Delta E$. The observed behavior
is not unexpected, as the reduced active spaces have been obtained
from a significant truncation of a much larger ``physical'' active space.
For the quantum simulations we chose to proceed considering the singlet states, which is the most stable. 
The QNP results in Table \ref{tab:QNP} for MOF and CO$_2$@MOF are based on 
minimal numbers of layers/parameters to achieve results within chemical accuracy; exact results are obtained for CO$_2$ using a 4 layer ansatz. 
Since these results are obtained from a variational approach, some degree of error cancellation can be expected when computing energy differences, such as in the case of the adsorption energy.
Indeed, the corresponding value for the adsorption energy (active space correlation contribution only, as computed from Eq. \ref{eq:Eadscorr})
is $\Delta E^{\mathrm{QNP}\mathrm{corr}}_{\mathrm{AS}}=25.3$ kJ/mol, which is in excellent agreement with the full CI reference in the reduced active space $\Delta E^{\mathrm{FCI}\mathrm{corr}}_{\mathrm{AS}}=25.4$ kJ/mol. This shows show that, at least in the ideal conditions of a classical simulator, the QNP ansatz provides satisfactory results 
also for relatively large systems involving around 30 qubits.

\begin{table*}[tb!]
\centering
\resizebox{0.7\textwidth}{!}{
\begin{tabular}{|l|l|l|l|l|}
\hline  \hline
System   & N qubits & N layers  &  N parameters &  Deviation from FCI (kJ/mol) \\
\hline \hline
CO$_2$ & 8 & 4 & 24 & $2.1\times 10^{-8}$ \\
MOF & 20 & 10 & 180 & 3.3 \\
CO$_2$@MOF & 28 & 18 & 468 & 3.4 \\
\hline \hline
\end{tabular}}
\caption{Hyperparameters used for the QNP ansatzes for the active spaces of CO$_2$, MOF, and
CO$_2$@MOF. The last column reports the deviations of the total energies with respect to the full CI reference.}.
\label{tab:QNP}
\end{table*}

\subsection{Quantum hardware experiments}

As shown in Sec. \ref{sec:QSresults}, the QNP ansatz performs reliably for the applications considered in this work and is employed here for initial state preparation in SQD. However, achieving chemical accuracy required 10 layers for MOF and 18 layers for CO$_2$@MOF. 
This entails quantum circuits with depths that are impractical for
current noisy hardware—an issue that becomes even more pronounced upon transpilation. In contrast, CO$_2$ easily converges with a minimal number of layers and, given the simplicity of its active space, it will not be discussed further in this section.

\begin{figure}[htbp]
  \centering
    \includegraphics[width=0.38\textwidth]{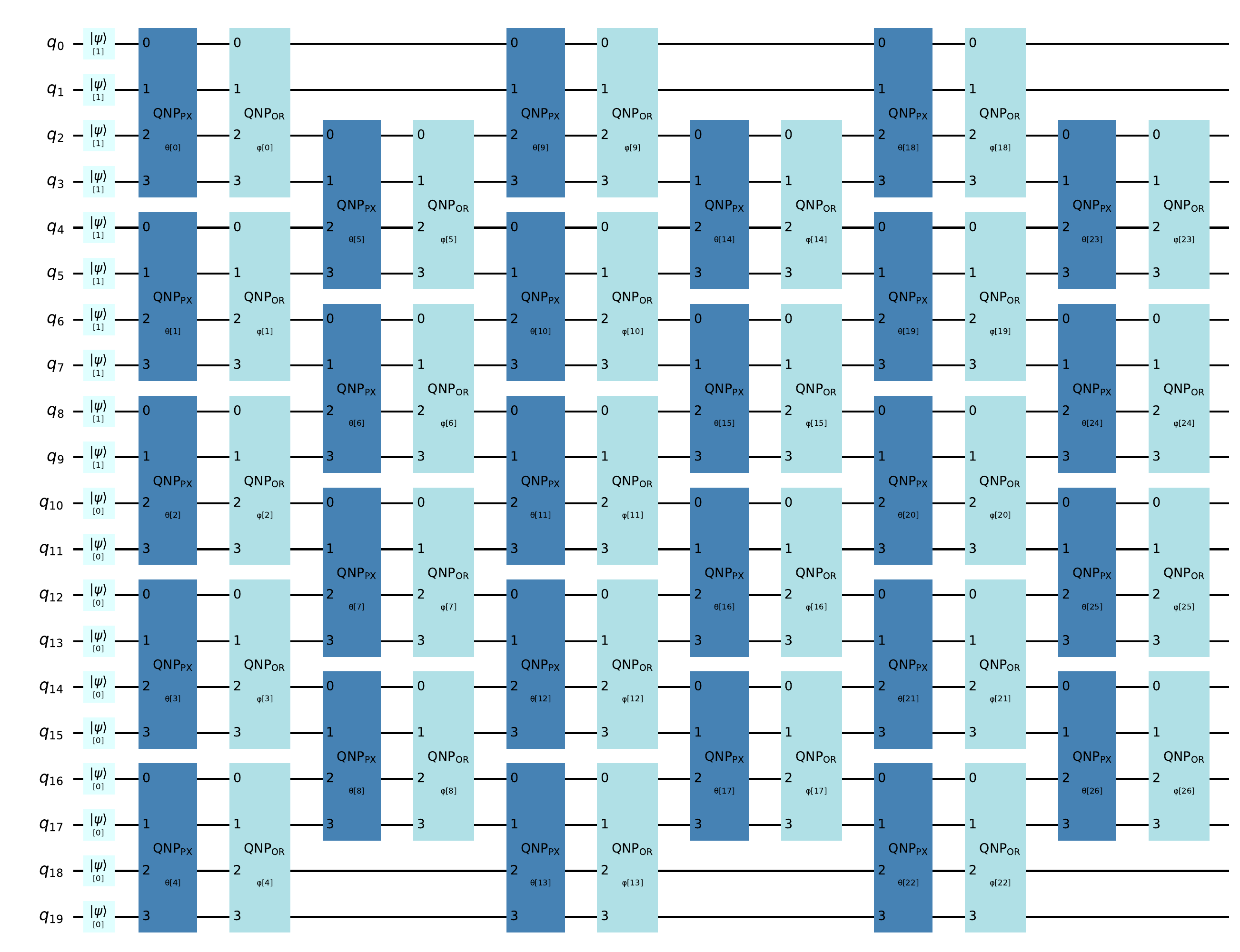}
    (a)\\[0.5cm]
    \includegraphics[width=0.34\textwidth]{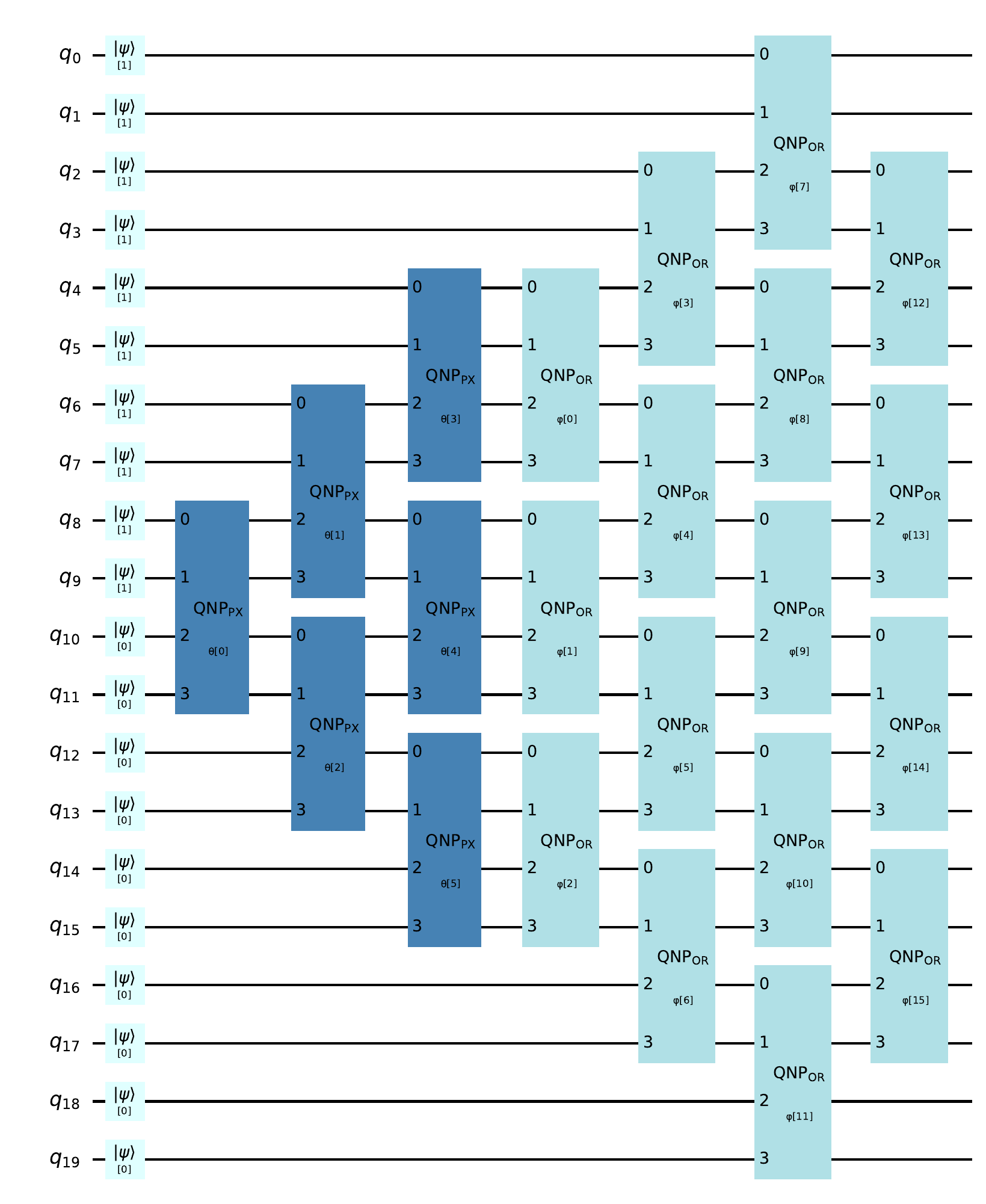}(b)
  \caption{Quantum circuit used to prepare the wavefunction $|\Psi\rangle$ in Eq. \ref{SQDgs} for the MOF active space. Starting from the QNP ansatz with three layers (a), a simplified circuit (b) suitable for quantum hardware implementation is obtained by removing gates with near-zero parameters or those that contribute only marginally to the energy.}
  \label{fig:reducedMOF}
\end{figure}

As discussed in Sec. \ref{sec:SQDmethod}, simpler ansatzes can be used in SQD applications, as the purpose is to obtain the correct configuration subspace rather that the exact wavefunction. 
To this purpose the QNP ansatz for MOF is truncated to 3 layers and the parameters are optimized using a classical simulator. Then gates that correspond to vanishing parameters or that do not contribute significantly to the final energy are removed, reducing the total number of QNP$_\mathrm{OR}$($\phi$) and QNP$_\mathrm{PX}$($\theta$) gates from 54 to just 22, as shown in Fig. \ref{fig:reducedMOF}. In practice, only an ``active cone'' of gates contribute to the ansatz. 
When used within VQE, both the full (Fig. \ref{fig:reducedMOF}(a)) and the truncated (Fig. \ref{fig:reducedMOF}(b)) quantum circuits are far from chemical accuracy, with deviations from full CI of 502 and 700 kJ/mol, respectively. As discussed below, these results are significantly improved by applying SQD.

The same approach is used to generate a shallower quantum circuit for CO$_2$@MOF but in this case 4 layers 
have been included in the QNP ansatz. By removing the gates with low contribution, the total number of QNP$_\mathrm{OR}$($\phi$) and QNP$_\mathrm{PX}$($\theta$) gates decreases from 104 in the full 4-layer QNP ansatz to just 40. At the VQE level the deviations of the energy from full CI are
605 and 644 kJ/mol for the full (Fig. \ref{fig:reducedMOF}(a)) and truncated (Fig. \ref{fig:reducedMOF}(b)) quantum circuits, respectively.

\begin{figure}[htbp]
  \centering
    \includegraphics[width=0.38\textwidth]{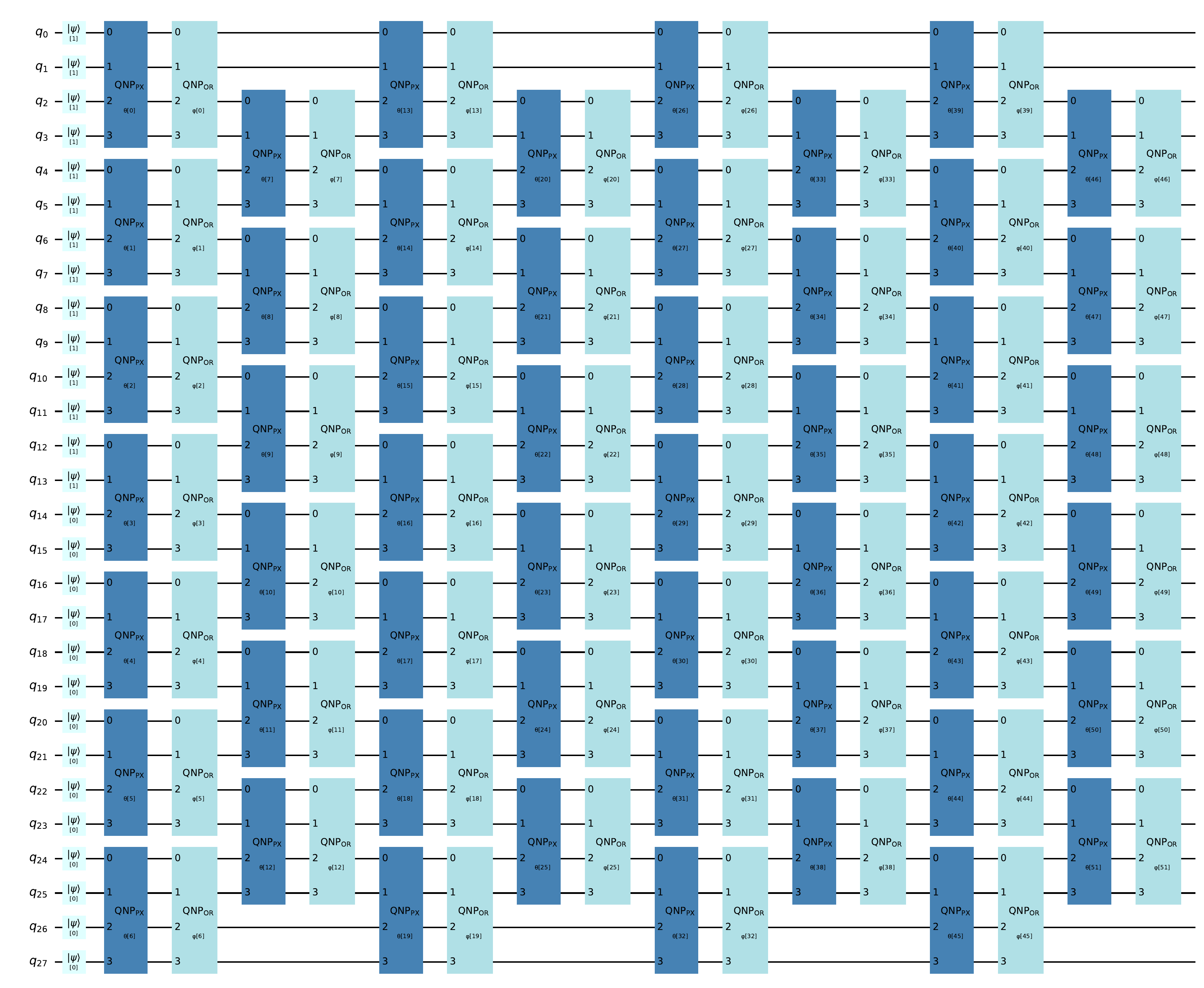}(a)\\[0.5cm]
    \includegraphics[width=0.34\textwidth]{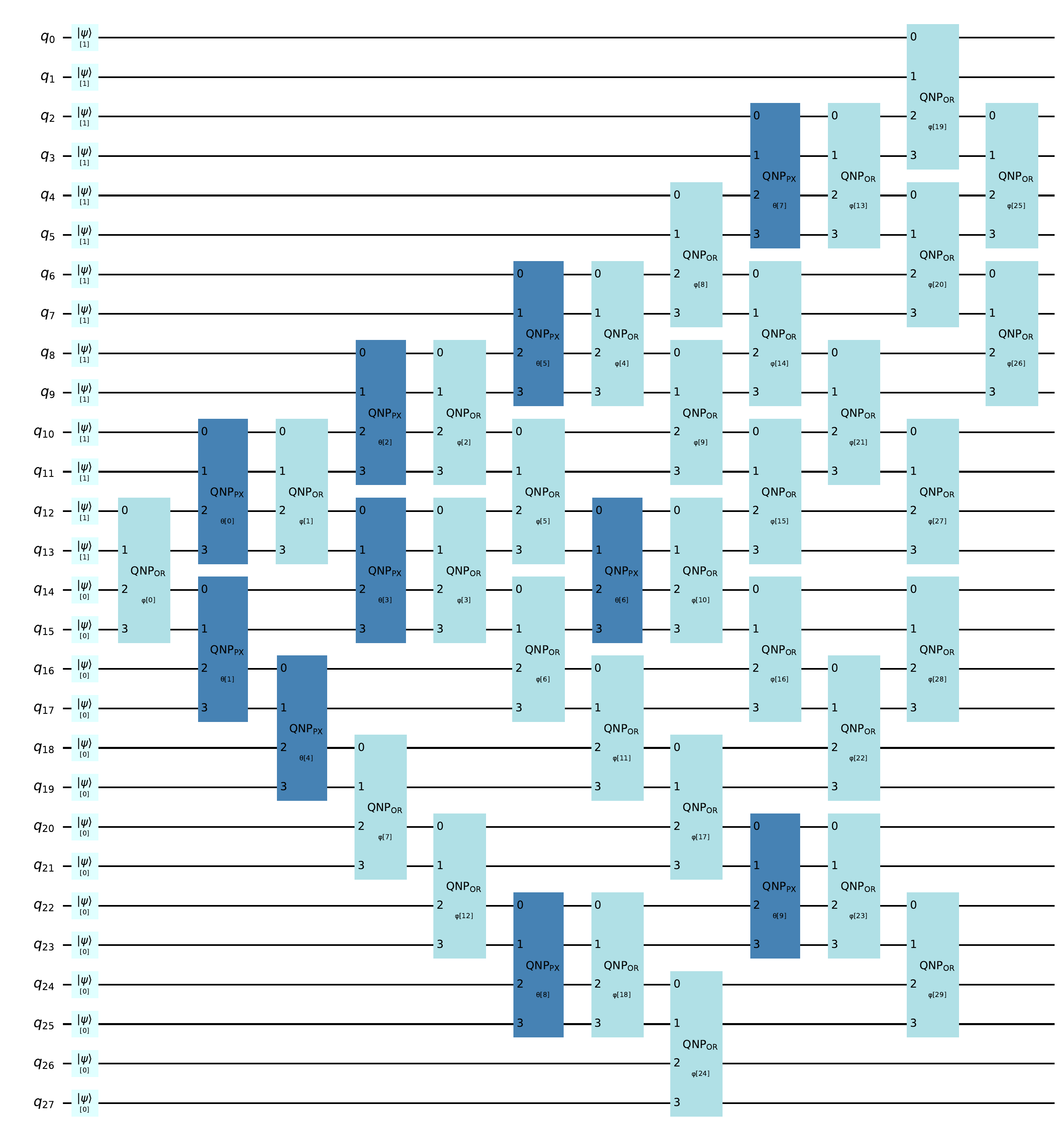}(b)
  \caption{Quantum circuit used to prepare the wavefunction $|\Psi\rangle$ in Eq. \ref{SQDgs} for the CO$_2$@MOF active space. Starting from the QNP ansatz with four layers (a), a simplified circuit (b) suitable for quantum hardware implementation is obtained by removing gates with near-zero parameters or those that contribute only marginally to the energy.}
  \label{fig:reducedMOFCO2}
\end{figure}

\begin{figure}[!t]
\centering
\includegraphics[width=0.8\columnwidth]{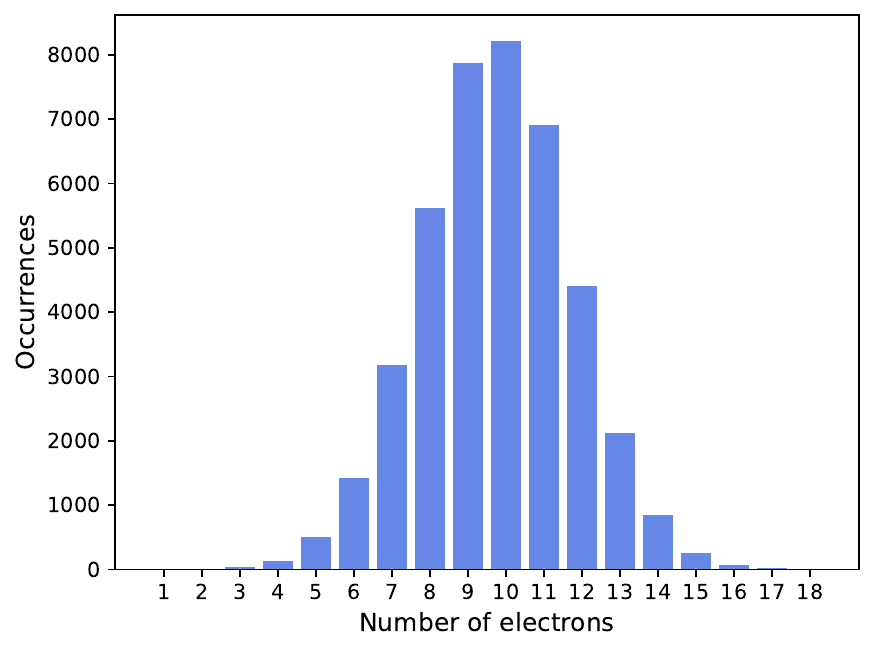}
\caption{Distribution of the hamming weights/number of electrons of the bit strings sampled in the quantum experiment for the MOF active space.}
\label{fig:OccurencesMOF}
\end{figure}

The reduced circuits have been used for   
experiments on the \texttt{ibm\_torino} quantum device. On this machine, the transpiled quantum circuits have a depth of 283 with 282 two-qubit CZ gates for MOF, and a depth of 405 with 592 two-qubit CZ gates for CO$_2$@MOF.
Using the SQD method, ground-state energies were estimated by sampling $10^5$ bit strings from the quantum circuits of both MOF and CO$_2$@MOF. Fig. \ref{fig:OccurencesMOF} shows the distribution of the Hamming weights (namely, the number of electrons) for the unique bit strings sampled. While the distribution peaks at the correct number of electrons (10), a substantial fraction of the sampled bit strings deviate and require correction. It is also important to notice that out of the 8207 configurations with 10 electrons, only 3958 have also the correct electron count in each spin channel. The behavior of the samples obtained for CO$_2$@MOF is analogous and will not be discussed further here.

The SQD ground-state energy results are summarized in Table \ref{tab:SQDenergies}, which reports deviations from the exact FCI values. 
The second column displays results obtained by postselecting only those configurations from the quantum experiment that possess the correct number of electrons in each spin channel. While these results are not yet chemically accurate, they are reasonably close to this threshold in the case of MOF. Upon applying the self-consistent configuration recovery of Ref. \citenum{robledo2024chemistry} to the full set of experimental samples, we observe a dramatic improvement in accuracy, that is approximately an order of magnitude beyond the threshold for chemical accuracy. This highlights the effectiveness of the configuration recovery technique, albeit at the expense of substantial classical computational effort.
For completeness, the table also includes, in parentheses, the corresponding $S^2$ values, which serve to confirm that the recovered ground states are indeed singlets.

\begin{table*}[htb!]
\centering
\resizebox{0.8\textwidth}{!}{
\begin{tabular}{|c|c|c|}
\hline  \hline
System       &
SQD energy   &  SQD energy \\ 
 & Postselection of correct configurations & Self-consistent recovery \\
 \hline \hline
MOF &  7.9 ($S^2$=0.015)  & 0.42 ($S^2$=0.00016) \\
CO$_2$@MOF & 77.5 ($S^2$=0.12)  & 0.28 ($S^2$=0.00015)  \\
\hline \hline
\end{tabular}}
\caption{Deviation of the ground-state energy (kJ/mol) computed with sample-based quantum diagonalization from the exact FCI reference. The configurations 
were obtained from hardware experiments by sampling the $\Psi$ states corresponding to the quantum circuits in Figs. \ref{fig:reducedMOF}(b) and \ref{fig:reducedMOFCO2}(b).}
\label{tab:SQDenergies}
\end{table*}

\section*{Conclusions}
In this work, we developed and demonstrated a quantum computing workflow for simulating CO$_2$ adsorption in periodic metal–organic frameworks, focusing on Fe-MOF-74 as a prototypical system. By combining plane-wave DFT calculations with an active space reduction strategy based on Wannier localization and MP2 natural orbitals, we achieved compact yet chemically meaningful orbital spaces suitable for near-term quantum algorithms. Using classical simulators, the quantum number-preserving (QNP) ansatz within the VQE framework was shown to reproduce full CI results within chemical accuracy in reduced active spaces. Furthermore, hardware experiments using sample-based quantum diagonalization (SQD) and self-consistent configuration recovery demonstrated that meaningful correlation energies can be obtained for active spaces up to 28 qubits, even on today’s noisy devices.

The workflow presented here provides a scalable path toward applying quantum computing to realistic surface science and adsorption problems in periodic materials. Future directions include extending this approach to larger and more physically complete active spaces, enabling the treatment of realistic spin multiplicities and yielding quantitatively accurate energy predictions. Pushing these limits will likely require not only more powerful quantum processors, but also improved error-mitigation strategies to preserve accuracy at increased circuit depths. These advances will be crucial for harnessing quantum computing in the predictive design of materials for carbon capture and related energy applications.

\section*{Author contributions}

Conceptualization, D.R., B.K., and H.W.L.; methodology, D.R., T.S., and A.G.; investigation, D.R., J.F.G., J.L., T.S., and A.G.; writing – original draft, D.R.; writing – review and editing, all authors; funding acquisition, B.K. and H.W.L.; supervision, D.R., B.K., and H.W.L.

\section*{Conflicts of interest}
There are no conflicts to declare.

\section*{Data availability}

The VASP code is copyrighted software and can be obtained from its official website.
Data are available from the corresponding authors upon reasonable request.

\section*{Acknowledgements}

This research was supported by Quantum Advantage challenge research based on Quantum Computing through the National Research Foundation of Korea (NRF) funded by the Ministry of Science and ICT (RS-2023-00257288).

The authors thank Robert Parrish (formerly at QC Ware Corp) for his valuable contributions during the earlier stages of this project.

\bibliographystyle{apsrev4-2}
\bibliography{MOF} 
\end{document}